\documentclass[12pt]{article}
\pdfoutput=1 
\usepackage{cmap}

\usepackage[X2,T3,T2A,T1]{fontenc}
\usepackage[utf8]{inputenc}  % newer: inputenx
\usepackage[ukrainian,russian.ancient,russian,english]{babel}  % last = primary language
\DeclareUnicodeCharacter{0462}{\CYRYAT}
\DeclareTextSymbolDefault{\CYRYAT}{X2}
\DeclareUnicodeCharacter{0463}{\cyryat}
\DeclareTextSymbolDefault{\cyryat}{X2}

\addto\captionsenglish{\renewcommand\refname{\normalsize Further references:}}
\usepackage{fancyhdr}
\makeatletter
\renewcommand{\@pnumwidth}{2.5em}% default is 1.55em
\makeatother

\newcounter{ppage}
\renewcommand{\theppage}{\mbox{P-{\arabic{ppage}}}}
\renewcommand{\thepage}{\theppage}
\renewcommand{\footnoterule}{
  \kern -3pt
  \hrule width 1.6cm
  \kern 2pt
}

\usepackage[dvipsnames]{xcolor}  

\usepackage{CJKutf8}
\newcommand{\zhs}[1]{\begin{CJK}{UTF8}{gbsn}#1\end{CJK}}

\usepackage{pdfpages,textcomp,multicol,tikz}
\usepackage{isoent,amssymb,revsymb}
\usepackage[pdfpagelabels=false,linktocpage=true,bookmarksnumbered=true,breaklinks=true,pagebackref=false,colorlinks=true,urlcolor=Mahogany,citecolor=Mahogany,linkcolor=Mahogany]{hyperref}
\usepackage{url}

\def\refname{\normalsize Further references:}

\begin{document}
\refstepcounter{ppage}
\parindent0.cm
\thispagestyle{empty}

\renewcommand{\thefootnote}{\fnsymbol{footnote}}

\begin{center}
\LARGE
Remarks on the empirical accuracy of the optical superposition 
principle [Engl.\ transl.\ of
Zh.\ Russk.\ Fiz.-Khim.\ Obshch., Ch.\ Fiz.\ {\bf 60}(1928)555]\\[1cm]

\large
Serge\u\i\ Ivanovich Vavilov 
[24.3.(O.S.\ 12.3.)1891-25.1.1951]\footnote[1]{\cite{1967poggendorf}, 
1987, {\it Teil [part] 8: Sn-Vl}, pp.\ 5655-5673.}

Institute of Biological Physics, Moscow

\end{center}

\renewcommand{\thefootnote}{\arabic{footnote}}

\vspace{1.5cm}

Full source details:

\vspace{3mm}

\begin{otherlanguage}{russian}
С.\ И.\ Вавилов
\end{otherlanguage}
%{\cyrrm S.\ I.\ Vavilov} 
[S.\ I.\ Vavilov]:
\begin{otherlanguage}{russian}
Замечания об эмпирической точности оптического принципа суперпозиции
\end{otherlanguage}
%{\cyrrm Zamechaniya ob \protect{\`{e}}mpirichesko\u i tochnosti
%opti\-cheskogo printsipa superpozitsii} 
[Zamechaniya ob \`empirichesko\u{\i} tochnosti
opticheskogo printsipa superpozitsii]/[Remarks on the 
empirical accuracy of the optical superposition principle].
{\it 
\begin{otherlanguage}{russian}
Журнал Русского Физико-Химического Общества при Ленинградском 
Университете, Часть Физическая
\end{otherlanguage}
%{\cyrrm
%\protect{Zh}urnal Russkogo Fizi\-ko-Khimicheskogo Obshchestva pri Leningradskom
%Universitete, Chast\cprime\ Fizicheskaya}
[Zhurnal Russkogo Fiziko-Khimicheskogo Obshchestva pri Leningradskom
Universitete, Chast'\ Fizi\-cheskaya]/ [Journal of the Russian 
Physico-Chemical Society at Leningrad University, Physical Part]}, 
Vol.\ LX--1928, No.\ 6, pp.\ 555-563. 
[in Russian, German summary]

Reprinted in (changed footnote numbering, German summary omitted):
\begin{otherlanguage}{russian}
С.\ И.\ Вавилов
\end{otherlanguage}
%{\cyrrm S.\ I.\ Va\-vilov} 
[S.\ I.\ Vavilov]:
{\it
\begin{otherlanguage}{russian}
Собрание сочинений. Том
\end{otherlanguage}
%{\cyrit Sobranie sochineni\u i. Tom} 
I - 
\begin{otherlanguage}{russian}
Работы по физике 1914-1936
\end{otherlanguage}
%{\cyrit Raboty po Fizike 1914-1936}
[Sobranie sochineni\u{\i}. Tom I - Raboty po fizike 
1914-1936]/[Collected Works. Vol.\ I - Works in Physics 1914-1936]}.
\begin{otherlanguage}{russian}
Издательство Академии наук СССР
\end{otherlanguage}
%{\cyrrm Izdatel\cprime stvo Akademii nauk SSSR}
[Izdatel'stvo Akademii nauk SSSR]/[Publishing House of the Academy of
Sciences of the USSR], Moscow, 1954, item 24, pp.\ 234-241
(The reprint is freely available online at the following site of the
Russian Academy of Sciences:
{\tt \url{http://nasledie.enip.ras.ru/ras/view/publication/general.html?id=43932907}}).

\vspace{0.7cm}

Translator:

\vspace{3mm}

K.\ Scharnhorst (E-mail: {\tt k.scharnhorst@vu.nl},\hfill\ \linebreak
ORCID: \url{http://orcid.org/0000-0003-3355-9663}),\hfill\ \linebreak
Vrije Universiteit Amsterdam,
Faculty of Sciences, Department of Physics and Astronomy,
De Boelelaan 1081, 1081 HV Amsterdam, The Netherlands

\newpage
\refstepcounter{ppage}

\ \\[4cm]

\tableofcontents

\newpage
\refstepcounter{ppage}

\phantomsection
\addcontentsline{toc}{section}{English translation of the article
{\rm (T-1 = P-3)}}

\renewcommand{\thepage}{\mbox{T-{\arabic{page}}}}
\chead{{\rm\color{gray}- \theppage\ -}\vspace{0.2cm}\linebreak
{\sf [Engl.\ transl.\ of Zh.\ Russk.\ Fiz.-Khim.\ Obshch., 
Ch.\ Fiz.\ {\bf 60}(1928)555-563]:\hfill - \thepage\ -}\linebreak\hrule}

\setcounter{page}{1}
\setcounter{footnote}{0}

\hspace{15.61cm}{\it 555}\nolinebreak\hspace{-16.43cm}

\begin{center}
{\LARGE\baselineskip0.9cm
Remarks on the empirical accuracy of the optical superposition 
principle.\footnote{Presented at the session of the optical section
of the VI Congress of Russian Physicists in Moscow on August 8, 1928}\\[1.5cm]}
{\large\it S.\ I.\ Vavilov.}\\[0.3cm]
\end{center}

\S\ 1. The following remarks concern the\ 
$\,$s$\,$u$\,$p$\,$e$\,$r$\,$p$\,$o$\,$s$\,$i$\,$t$\,$i$\,$o$\,$n$\,$\ 
o$\,$f\ $\,$t$\,$h$\,$e\ 
$\,$i$\,$n$\,$t$\,$e$\,$n$\,$s$\,$i\-$\,$t$\,$i$\,$e$\,$s\ 
$\,$o$\,$f\ $\,$i$\,$n$\,$c$\,$o$\,$h$\,$e$\,$r$\,$e$\,$n$\,$t\ 
$\,$b$\,$u$\,$n$\,$d$\,$l$\,$e$\,$s\ $\,$o$\,$f\ 
$\,$l$\,$i$\,$g$\,$h$\,$t. In the theory of 
wave motion this principle is being considered as an average statistical
sequel of the addition theorem of amplitudes whereby the latter theorem 
is either completely exact (electromagnetic theory), or has a limited 
validity for small amplitudes (elastic theories). In the following, the 
superposition principle is mainly considered as an\ 
$\,$e$\,$x$\,$p$\,$e$\,$r$\,$i$\,$m$\,$e$\,$n$\,$t$\,$a$\,$l\ 
$\,$f$\,$a$\,$c$\,$t$\,$\ which shall play a certain role in making judgements 
on the soundness of one or the other conception of the nature of light.
$\,$H$\,$u$\,$y$\,$g$\,$h$\,$e$\,$n$\,$s\footnote{\label{555footnote2}C$\,$h.$\,$ 
\ H$\,$u$\,$y$\,$g$\,$e$\,$n$\,$s, 
Abhandlung \"uber das Licht. (1678) Ostw.\ Kl.\ \textnumero\ 20, p.\ 25.}
wrote on this empirical principle: 
``Another property of waves of light, and one of the most marvellous,
is that when some of them come from different or even from opposing
sides, they produce their effect across one another without any
hindrance.''.

It is being said that the independence of the propagation 
of a bundle of light of the passage of
other bundles on its path is explained by the wave 
theory of light, however, this is not so easily reconciled with 
the primitive corpuscular idea. The reproach often made in the XVIII.\
century to the Newtonian theory 
consisted just in saying that the collisions of the light
corpuscles, i.e., a violation of the superposition,
should be observed.
An answer to this difficulty was the admission of the extreme
smallness of the corpuscles: ``I know
-- $\,$L$\,$o$\,$m$\,$o$\,$n$\,$o$\,$s$\,$o$\,$v\footnote{
\label{555footnote3}
\begin{otherlanguage}{russian}
М.\ В.\ Ломоносов, Слово о происхождении света. Собрание разных 
сочинений, ч.\
\end{otherlanguage}
%{\cyrrm 
%M.\ V.\ Lomonosov, Slovo o proiskhozhdenii sveta. Sobranie raznykh 
%sochineni\u i, ch.}\
III
[M.\ V.\ Lomonosov, Slovo o proiskhozhdenii sveta. Sobranie raznykh 
sochineni\u{\i}, ch.\ III], 1803, p.\ 155.}
wrote addressing the defenders of the corpuscular hypothesis
-- that you divide 
the material of light into such fine 
particles and place them in universal space with so little
density that the whole quantity can be compressed and packed
in the porous crevices of one grain of sand.''. 
Presently, optics is in the stage of the theoretical synthesis
of the wave and corpuscular conceptions and the problem of the 
limits of the realizability of the superposition is becoming 
somewhat unclear again. Is it possible to assign to a 
quantum centre, accompanied by a wave, a reality of that
kind that in encounters,

\hspace{15.61cm}{\it 556}\nolinebreak\hspace{-16.43cm}
``collisions'' of light quanta a perturbation of the initial
propagation occurs?

If a violation of the superposition of the\
$\,$i$\,$n$\,$t$\,$e$\,$n$\,$s$\,$i$\,$t$\,$i$\,$e$\,$s$\,$\ of 
light bundles is possible at all, 
according to the principle of energy conservation 
it can only make its appearance by a change in the direction of light,
its\ $\,$s$\,$c$\,$a$\,$t$\,$t$\,$e$\,$r$\,$i$\,$n$\,$g, 
or a peculiar diffraction of waves off other, incoherent light quanta.

Experiment and theory have established a deep analogy between material
particles and light quanta, both are accompanied by waves and 
the energy is concentrated
in some kernel. In encounters of atomic or electron beams as well as
in encounters of light and matter a scattering or diffraction of
waves occurs. The scattering of electrons and atoms can qualitatively be
explained by the occurrence of electrostatic forces independently
of the wave theory of matter. The scattering of light by electrons 
and atoms has a wave as well as a corpuscular interpretation, however,
the descriptions of the two partners in such a scattering are not the 
same. For the material partner the kernel is well studied and its 
waves are less known, conversely the data for the light waves are
very precise but the information over the quantum kernel is limited
to its energy and momentum. The study of the results of encounters
of light quanta should clarify to some extent the questions 
arising here: Does in such an encounter a complete unhindered mutual
penetration occur, or has a peculiar ``impenetrability'' to be ascribed to
light quanta, and is, consequently, a ``selfscattering'' of light, 
or a diffraction of light waves off light quanta, possible? Can the 
quanta any effective size, i.e., a violation of the superposition 
in encounters be ascribed to as it is being 
done for atoms and protons on the basis of scattering?

Many times in recent years, attempts have been made to hypothetically 
ascribe to light quanta linear dimensions of the order of the
wavelength of light waves. If these dimensions aspire to have some 
physical reality they could manifest themselves by the
``selfscattering'' mentioned. At the present stage of the theory
of light, the question of the ``extension'' of the quanta, or, more
precisely of the possible deviations from the superposition apparently
cannot be considered as conclusively solved in the positive or negative 
sense and, therefore, one has primarily to turn to the experiment.

\refstepcounter{ppage}
\S\ 2. A noticeable violation of the superposition, if possible
at all, can obviously be expected at very large radiation energy
densities only. As can be calculated easily, if the effective size 
of the quanta was of the order of the light wave we should observe 
a completely noticeable scattering of light when two 
bundles from the most modest light sources intersect. This is not so
and, therefore, in reality the magnitude of such a quantity in any case 
has to be physically small or has a formal, symbolic character.

\hspace{15.61cm}{\it 557}\nolinebreak\hspace{-16.43cm}
Under laboratory conditions, the highest\  
$\,$i$\,$n$\,$s$\,$t$\,$a$\,$n$\,$t$\,$a$\,$n$\,$e$\,$o$\,$u$\,$s$\,$\ 
radiation densities can perhaps be obtained 
by means of the light of a condensed spark. 
Concentrating this light with a lense,
instantaneous values of the light energy density 
that exceed the corresponding value on the surface of the sun
can be achieved easily. 
In this case, the average density is small due to the short duration and
rareness of the sparks, but the hypothetical effect of the ``collisions''
of light quanta must be proportional to the\  
$\,$s$\,$q$\,$u$\,$a$\,$r$\,$e$\,$\ of 
the instantaneous density, therefore, a spark turns out to be 
considerably more advantageous than, for example, an arc. 
Preliminary experiments with a spark the light of which met in an
evacuated container\
$\,$d$\,$i$\,$d\ $\,$n$\,$o$\,$t\
$\,$u$\,$n$\,$c$\,$o$\,$v$\,$e$\,$r\ $\,$a$\,$n$\,$y\ 
$\,$n$\,$o$\,$t$\,$i$\,$c$\,$e$\,$a$\,$b$\,$l$\,$e\
$\,$s$\,$c$\,$a$\,$t$\,$t$\,$e$\,$r$\,$i$\,$n$\,$g.
These observations have been carried out with the usual
precautions, in front of distant container walls covered with 
black velvet; for control, the experiments have been repeated
with the light of an incandescent lamp that delivered the same\ 
$\,$a$\,$v$\,$e$\,$r$\,$a$\,$g$\,$e$\,$\ radiation 
density; in both cases the result has been equally negative.

Incidentally, laboratory conditions are considerably surpassed 
in this respect by what observations of the sun yield. At the surface of the 
sun incoherent bundles originating from different radiating sections
are intersecting each other; 
the intersections occur at very large radiation densities and
in a huge volume whereby for an observer on earth
the results sum up. In moments of total solar eclipse, 
when direct rays are blocked
and the background is very dark we meet exceptionally 
good observational conditions, and the sun serves as the most convenient
object for the establishment of the limits of 
the satisfiability of the superposition.

\setcounter{footnote}{0}
This way, the considered problem immediately makes contact with the problem
of the\ $\,$s$\,$o$\,$l$\,$a$\,$r\
$\,$c$\,$o$\,$r$\,$o$\,$n$\,$a. Until now, the nature of the corona remains
mysterious in general and in particular. \footnote{A critical analysis of
the different theories of the corona is given in the works of\ 
$\,$A$\,$n$\,$d$\,$$\,$e$\,$r$\,$s$\,$o$\,$n:\ 
$\,$W$\,$i$\,$l$\,$h$\,$e$\,$l$\,$m\ $\,$A$\,$n$\,$d$\,$e$\,$r$\,$s$\,$o$\,$n.
ZS.\ f.\ Phys.\ {\bf 28}, 299, 1924; {\bf 33}, 273, 1925;
{\bf 34}, 453, 1925; {\bf 35}, 757, 1925; {\bf 37}, 342, 1926.} 
The existing explanations for the corona
as scattered light from cosmic dust surrounding the sun, as scattering 
from the molecules of a diluted gas, or an electron
atmosphere, as fluorescence of a gas etc.\ are undoubtedly partially 
correct; however, neither taken for themselves nor together
these factors are capable of quantitatively explaining the most
intense part of the emission of the corona. In particular, for the
explanation of the corona by means of the scattering in an electron
gas one has to make hardly justified assumptions over the violation 
of the laws of electrodynamics near the sun.

\refstepcounter{ppage}
In any case, for an observer on earth the corona\
$\,$i$\,$s\ $\,$a$\,$n\ 
$\,$i$\,$n$\,$e$\,$s$\,$c$\,$a$\,$p$\,$a$\,$b$\,$l$\,$e\ 
$\,$c$\,$o$\,$n$\,$d$\,$i$\,$\-t$\,$i$\,$o$\,$n\ 
$\,$f$\,$o$\,$r\ $\,$t$\,$h$\,$e\ 
$\,$o$\,$b$\,$s$\,$e$\,$r$\,$v$\,$a$\,$t$\,$i$\,$o$\,$n$\,$\ 
which places a certain empirical 
accuracy limit on the assertion concerning the correctness
of the superposition principle. Undoubtedly, at least
a part of the general emission of the corona consists of light scattered off
matter and, therefore, has no relation to a possible violation
of the superposition.

\setcounter{footnote}{0}
\hspace{15.61cm}{\it 558}\nolinebreak\hspace{-16.43cm}
However, the general emission allows to establish an\ 
$\,$u$\,$p$\,$p$\,$e$\,$r\ $\,$b$\,$o$\,$u$\,$n$\,$d$\,$\
on the correctness of the principle.

For an approximate estimate of this bound one can rely on the fact
that half of the total emission of the corona, as photographs show, is 
distributed within a distance of half of the solar radius taken from
the edge of the sun.
For a more accurate calculation one can make use of the empirical 
formulas for the decay law of the coronal intensity with distance.
For example, according to data for the eclipse of June 29, 1927 by\  
$\,$B$\,$a$\,$l$\,$a$\,$n$\,$o$\,$v$\,$s$\,$k$\,$y$\,$\ and\
$\,$P$\,$e$\,$r$\,$e$\,$p$\,$e$\,$l$\,$k$\,$i$\,$n\ 
\footnote{\label{558footnote1}I.\ 
$\,$B$\,$a$\,$l$\,$a$\,$n$\,$o$\,$v$\,$s$\,$k$\,$y$\,$\ and\ 
$\,$E.\ $\,$P$\,$e$\,$r$\,$e$\,$p$\,$e$\,$l$\,$k$\,$i$\,$n.
Monthly Notices of. Roy.\ Astr.\ Soc.\  {\bf 88}, 740, 1928.}
the coronal intensity $I$ at distance
$R$ from the center of the sun ($R$ -- expressed in terms of
solar radii) is given by the formula:
\begin{equation}
\label{vav1}
I\ =\ \frac{A_1}{(R - 0,849)^3}\ \ .\ \ .\ \ .\ \ .\ \ .\ \ 
.\ \ .\ \ .\ \ .\ \ .\ \ .\ \ .\ \ .\ \ .\ \ .\ \ .\ \ .\ \ .
\end{equation}

From this, it can easily be calculated that one half of the general 
emission of the corona is concentrated within the range of 0,4 of the solar 
radius. The total emission of the corona approximately amounts to
$1\cdot 10^{-6}$ of the solar radiation. \footnote{Cf.\ 
$\,$M$\,$\"u$\,$l$\,$l$\,$e$\,$r$\,$-$\,$$\,$Po$\,$u$\,$i$\,$l$\,$l$\,$e$\,$t, 
Lehrbuch der Physik, B. V, 2-nd part,
p.\ 157.}
One half of this amount, i.e.\ $5\cdot 10^{-7}$, is being scattered
within a distance of 0,4 of the solar radius, i.e.\ on a path of 
$2,8\cdot 10^{10}$ {\it cm}. This way, on average on each {\it cm}
of its path near the sun the light ray loses\
$\,$d$\,$u$\,$e\ $\,$t$\,$o\ $\,$a$\,$l$\,$l\ 
$\,$s$\,$o$\,$u$\,$r$\,$c$\,$e$\,$s\ $\,$o$\,$f\ 
$\,$s$\,$c$\,$a$\,$t$\,$t$\,$e$\,$r$\,$i$\,$n$\,$g\
$\,$e$\,$x$\,$i$\,$s$\,$t$\,$e$\,$n$\,$t\ 
$\,$t$\,$h$\,$e$\,$r$\,$e$\,$\ 
$\frac{\displaystyle 5\cdot 10^{-7}}{\displaystyle 2,8\cdot 10^{10}}$\ , 
i.e.\ roughly
\begin{center}
\fbox{\parbox{1.8cm}{$1,8\cdot 10^{-17}$}}
\end{center}
\refstepcounter{ppage}
of its energy. This is the upper bound on the accurary level 
of the superposition principle near the sun, i.e.,
under exceptionally favourable conditions for its assumed violation.
Even in the absence of any competing sources for the 
``selfscattering'' of light one could associate $1,8\cdot 10^{-17}$ 
of the energy to intersecting, very intense light bundles only. 
From this it is clear how difficult if not hopeless in general it is
to observe such a scattering in the laboratory, even if it would be real.
With a condensed spark
one can easily achieve within a short time of roughly 
$1\cdot 10^{-6}$ {\it sec.}\ a light flux with an intensity of
$10^{11}\ \frac{\displaystyle\it erg.}{\displaystyle\it cm^2} {\it sec.}$,
such as near the sun. In the most favourable case, 
the light scattering per second\
$\,$i$\,$n$\,$t$\,$o\ $\,$a$\,$l$\,$l\
$\,$d$\,$i$\,$r$\,$e$\,$c$\,$t$\,$i$\,$o$\,$n$\,$s$\,$\ 
could amount to (at 100 sparks per second)
$1,8\cdot 10^{-17}\cdot 10^{11}\cdot10^{-6}\cdot 10^2 =  1,8\cdot 10^{-10}$
ergs per second. Practically, a quantity tens or even hundreds times
smaller than the visual stimulus threshold would have to be observed.

\setcounter{footnote}{0}
\hspace{15.61cm}{\it 559}\nolinebreak\hspace{-16.43cm}
This way, the weakness of the emission of the solar
corona shows that the superposition principle, at least
for visible light, is one of the most precise fundamentals of optics.

\S\ 3. Is it possible to firmly establish such wide limits for the 
applicability of the superposition and at the same time to see 
in the solar corona an argument in favour of the existence of the 
``selfscattering of light''? If some astrophysical theory suceeds in 
giving a convincing explanation for the most intense part of the
corona the question of course disappears; as long as it does not exist
and the theory of light has not yet assumed any concrete form,
the hypothesis of the ``selfscattering'' has apparently
the right to be discussed on an equal footing with other
proposals. As has been explained above,
unfortunately a direct manner of resolution, an
experimentum crucis on earth, can hardly be realized.

Certain characteristic features of the corona find a simple 
explanation from the viewpoint of the ``selfscattering'' hypothesis. 
The coronal intensity quickly decays with distance from the edge
of the sun. Some empirical formulas expressing the 
dependence of the intensity on the distance exist, 
\footnote{Cf.\ W.\ 
$\,$A$\,$n$\,$d$\,$e$\,$r$\,$s$\,$o$\,$n. ZS.\ f.\ Physik {\bf 35}, 757, 1926.}
these formulas strongly differ from each other. For small distances
from the sun, in some cases the formula by\ 
$\,$T$\,$u$\,$r$\,$n$\,$e$\,$r$\,$
\footnote{\label{559footnote2}Cf.\ E.\ 
$\,$P$\,$e$\,$t$\,$t$\,$i$\,$t$\,$\  and\
$\,$S$\,$e$\,$t$\,$h.\ $\,$B.\ 
$\,$N$\,$i$\,$c$\,$h$\,$o$\,$l$\,$s$\,$o$\,$n. 
Astrophys.\ Journ.\ {\bf 62}, 202, 1925.}
is well fulfilled:
\begin{equation}
\label{vav2}
I\ =\ \frac{A_2}{R^6}\ \ .\ \ .\ \ .\ \ .\ \ .\ \ 
.\ \ .\ \ .\ \ .\ \ .\ \ .\ \ .\ \ .\ \ .\ \ .\ \ .\ \ .\ \ .
\end{equation}
where $R = \frac{\displaystyle r}{\displaystyle r_0}$
($r$ -- distance from the center of the sun, 
$r_0$ -- solar radius). B$\,$e$\,$c$\,$k$\,$e$\,$r\ \ 
proposed the formula:
\begin{equation}
\label{vav3}
I\ =\ \frac{A_2}{(R-0,72)^4}\ \ .\ \ .\ \ .\ \ .\ \ .\ \ 
.\ \ .\ \ .\ \ .\ \ .\ \ .\ \ .\ \ .\ \ .\ \ .\ \ .\ \ .\ \ .
\end{equation}

The cubic formula (\ref{vav1}) has been given above. For large distances,
the results of observations sometimes agree well with the formula
by\ $\,$B$\,$e$\,$r$\,$g$\,$s$\,$t$\,$r$\,$a$\,$n$\,$d:
\begin{equation}
\label{vav4}
I\ =\ \frac{A_2}{(R-1)^2}\ \ .\ \ .\ \ .\ \ .\ \ .\ \ 
.\ \ .\ \ .\ \ .\ \ .\ \ .\ \ .\ \ .\ \ .\ \ .\ \ .\ \ .\ \ .
\end{equation}

Apparently, until now it has not been clarified if such a variety of
formulas corresponds to real changes in the corona or is a result
of the imperfection of the measurements.

From the viewpoint of the hypothesis of light quanta collisions,
the intensity of the scattered light at a given point should be proportional 
to the number of the collisions of quanta, i.e., primarily it should be
proportional
to the product of the\ $\,$s$\,$q$\,$u$\,$a$\,$r$\,$e$\,$\ of the 
radiation density and the relative velocity of the colliding quanta.
\linebreak

\vspace{-0.55cm}

\refstepcounter{ppage}
\setcounter{footnote}{0}
\hspace{15.61cm}{\it 560}\nolinebreak\hspace{-16.43cm}
Let the line $ED$ on fig.\ 1 connect a point under consideration $A$ near the 
sun with the earth $E$. We denote by $r_0$ -- the radius of the sun,
by $r$ -- the distance $CG$ (the perpendicular from the center of the
sun onto the line $ED$), by $x$ -- the distance $AG$. The radiation 
density in the point $A$ is proportional to: \footnote{Cf.\ W.\ 
$\,$A$\,$n$\,$d$\,$e$\,$r$\,$s$\,$o$\,$n, loc.\ cit.} 
\begin{eqnarray*}
&1\ -\ \sqrt{1 - \frac{\displaystyle r_0^2}{\displaystyle x^2 + r^2}}& .
\end{eqnarray*}

The average relative velocity of quanta colliding in the point $A$
is proportional to $\sin$ of the angle formed by the\ 
$\,$a$\,$v$\,$e$\,$r$\,$a$\,$g$\,$e\ bundle
of the cone of rays meeting in $A$ and the axis of the cone.

Therefore, the average relative velocity of the quanta colliding in
the point $A$ is proportional to:
\begin{eqnarray*}
&\sqrt{\frac{\displaystyle r_0^2}{\displaystyle x^2 + r^2}}& .
\end{eqnarray*}

\begin{multicols}{2}
\

\begin{tikzpicture}(140,140)
%\draw [help lines] (0,0) grid (7,5);
\draw (4.694,2.353) circle (1.06cm);
\draw[dashed] (4.694,2.353) circle (2.31cm);
\draw (0.353,2.4)--(5.894,4.741);
\draw[dashed] (0.353,2.4)--(5.729,4.388);
\draw (4.694,2.353)--(4.024,3.95);
\draw (4.694,2.353)--(3.012,3.52);
\draw (4.295,1.371)--(5.093,1.371);
\draw (4.139,1.45)--(5.249,1.45);
\draw (4.027,1.529)--(5.361,1.529);
\draw (3.941,1.607)--(5.448,1.607);
\draw (3.87,1.686)--(5.518,1.686);
\draw (3.813,1.764)--(5.575,1.764);
\draw (3.765,1.843)--(5.623,1.843);
\draw (3.726,1.921)--(5.662,1.921);
\draw (3.694,2)--(5.694,2);
\draw (3.67,2.079)--(5.718,2.079);
\draw (3.652,2.157)--(5.736,2.157);
\draw (3.641,2.236)--(5.748,2.236);
\draw (3.635,2.314)--(5.753,2.314);
\draw (3.635,2.393)--(5.753,2.393);
\draw (3.641,2.471)--(5.747,2.471);
\draw (3.6525,2.55)--(5.736,2.55);
\draw (3.67,2.628)--(5.718,2.628);
\draw (3.695,2.707)--(5.693,2.707);
\draw (3.726,2.785)--(5.662,2.785);
\draw (3.765,2.864)--(5.623,2.864);
\draw (3.815,2.943)--(5.573,2.943);
\draw (3.871,3.021)--(5.517,3.021);
\draw (3.942,3.1)--(5.446,3.1);
\draw (4.028,3.178)--(5.36,3.178);
\draw (4.14,3.257)--(5.248,3.257);
\draw (4.295,3.335)--(5.093,3.335);
\draw (3.05,3.8) node {\sf A};
\draw (4.9,2.35) node {\sf C};
\draw (5.55,4.8) node {\sf D};
\draw (0.425,2.7) node {\sf E};
\draw (4,4.175) node {\sf G};
\end{tikzpicture}

\hspace{2cm}Fig.\ 1. \\[5cm]

\refstepcounter{ppage}

The rate and direction of the scattering of light quanta can turn out
to depend on the angle between the colliding quanta. Nothing can 
be said about this dependence because the phenomenon of ``collisions'' 
of quanta itself is completely hypothetical and unspecified. Formally,
we characterize this dependence by means of a certain function
\begin{eqnarray*}
&f(r,x).&
\end{eqnarray*}
\end{multicols}

The scattering from all volume elements lying infinitesimally close
to the solid angle 
related to the vertex on earth merges for an observer on earth
in perspective into one point of the corona; the number of 
volume elements growths proportionally to the square distance from earth,
in turn the scattering intensity reaching the earth from these
volumes decays inversely proportional with the square distance; this
way, the distance from earth does not enter into the expression 
for the total scattering. In accordance with the assumptions made,
the emission at a certain point of the corona at distance $r$ 
from the center of the sun is expressed by the formula:
\begin{equation}
\label{vav5}
I\ =\ A_5\ \int\limits_0^\infty
\left[1\ -\ \sqrt{1 - \frac{r_0^2}{x^2 + r^2}}\ \right]^2
\cdot
\left[\sqrt{\frac{r_0^2}{x^2 + r^2}}\ \right]\cdot f(r,x)
\cdot dx.\ \ .\ \ .
\end{equation}

Here, the assumption has been made that the radiation from all
elements of the\linebreak

\vspace{-0.55cm}

\refstepcounter{ppage}
\setcounter{footnote}{0}
\hspace{15.61cm}{\it 561}\nolinebreak\hspace{-16.43cm}
sun's surface is equal (this, of course, is not 
exactly the case) and should read
\begin{equation}
\label{vav6}
A_5\ =\ k E^2,\ \ .\ \ .\ \ .\ \ .\ \ .\ \ 
.\ \ .\ \ .\ \ .\ \ .\ \ .\ \ .\ \ .\ \ .\ \ .\ \ .\ \ .\ \ .
\end{equation}
where $E$ -- is the solar constant.

We assume that $f(r,x)$ does not depend, or weakly depends, on $r$.

Furthermore, we restrict ourselves to distances from the edge of
the sun at which the average value of $x^2 + r^2$ under the integral
is considerably larger than $r_0^2$. Under these conditions the
equation (\ref{vav5}) assumes the simple form:
\begin{equation}
\label{vav7}
I\ =\ \frac{A_6}{R^4},\ \ .\ \ .\ \ .\ \ .\ \ .\ \ 
.\ \ .\ \ .\ \ .\ \ .\ \ .\ \ .\ \ .\ \ .\ \ .\ \ .\ \ .\ \ .
\end{equation}
where $R$ has the same meaning as before.

An exact comparison of the empirical formulas 
(\ref{vav1}), (\ref{vav2}), (\ref{vav3}),
(\ref{vav4}) with the theoretical formulas (\ref{vav6}) 
and (\ref{vav7}) makes no sense.
The empirical formulas do not agree with each other, on the 
other hand a certain part of the emission of the corona is
without any doubt caused by light scattering in matter and, 
therefore, the theoretical 
formula cannot fully be applied to the corona. Under these
provisions, 

\begin{center}
T$\,$A$\,$B$\,$L$\,$E$\,$\ 1.

\vspace{2mm}

\begin{tabular}{|c|c|c|c|c|c|}\hline\hline
& & & & & \\[-2mm]
R & (\ref{vav1}) & (\ref{vav2}) & (\ref{vav3}) & (\ref{vav4}) & (\ref{vav7})\\
& & & & &\\[-2mm]
\hline\hline
& & & & &\\
1,8 & 11,6\ \ & 21,0\ \ & 20,0\ \ & \ 6,1\ \ & 7,7\ \ \\
2,0 & \ 6,5\ \ & 10,4\ \ & 10,0\ \ & 3,9\ \ & 5,0\ \ \\
2,5 & \ 2,2\ \ & \ 2,9\ \ & \ 2,7\ \ & 1,8\ \ & 2,0\ \ \\
3,0 & \ 1,0\ \ & \ 1,0\ \ & \ 1,0\ \  & 1,0\ \ & 1,0\ \ \\
3,5 & \ 0,55 &  \ 0,40 & \ 0,45 & 0,64 & 0,54\\
4,0 & \ 0,31 &  \ 0,18 & \ 0,24 & 0,43 & 0,31\\
4,5 & \ 0,20 &  \ 0,09 & \ 0,14 & 0,32 & 0,20\\
5,0 & \ 0,14 &  \ 0,05 & \ 0,18 & 0,25 & 0,13\\[-2mm]
& & & & &
\end{tabular}
\end{center}

however, one may speak of a satisfactory agreement between 
the theoretical and experimental formulas. In Table 1, the 
relative values for the intensity
of the corona from $R = 1,8$
to  $R = 5,0$, calculated by means of the formulas (\ref{vav1}), (\ref{vav2}),
(\ref{vav3}), (\ref{vav4}), and (\ref{vav7}), 
are listed, where the values for  $R = 3,0$
are taken to be equal to 1. From the table, it can be seen that
formula (\ref{vav7}) well agrees with the most recent empirical 
formula (\ref{vav1}).
The agreement in the upper figures will increase if one does 
not confine oneself to the approximate formula (\ref{vav7}), but calculates
the values $I$ according to the fundamental formula (\ref{vav5}) by
assuming $f(x,r)$ to be constant. An acceptable agreement also
exists with formula (\ref{vav4}), on the other hand the formulas of\
$\,$T$\,$u$\,$r$\,$n$\,$e$\,$r$\,$\ (\ref{vav2}) and\
$\,$B$\,$e$\,$c$\,$k$\,$e$\,$r$\,$\ (\ref{vav3}) sharply disagree with 
(\ref{vav7}).

\refstepcounter{ppage}
\setcounter{footnote}{0}
\hspace{15.61cm}{\it 562}\nolinebreak\hspace{-16.43cm}
As is clear from above, in the formulas (\ref{vav5}) and (\ref{vav7}) the 
constants $A_5$ or $A_6$ are proportional to the product of the intensities
of the colliding bundles. If the radiation from the surface of the 
sun would be allover equal, $A_5$ and $A_6$ would simply be 
proportional to the\ $\,$s$\,$q$\,$u$\,$a$\,$r$\,$e$\,$\ of the solar 
constant, what is also put as basis of (\ref{vav5}).

Due to the inhomogeneity of the emission of the sun the values 
of the constants $A_5$ and $A_6$ should fairly strongly differ for
different radial directions of the corona, staying, however,
practically unchanged along a given radius what agrees with 
observations\footnote{Cf.\ H.\ 
$\,$L$\,$u$\,$d$\,$e$\,$n$\,$d$\,$o$\,$r$\,$f$\,$f, Sitzb.\ d.\
preuss.\ Akc.\ d.\ W., V, 1925, also. I.\ 
$\,$B$\,$a$\,$l$\,$a$\,$n$\,$o$\,$v$\,$s$\,$k$\,$y$\,$\  
and\ $\,$E.\ $\,$P$\,$e$\,$r$\,$e$\,$p$\,$e$\,$l$\,$k$\,$i$\,$n, loc.\ cit.}.
From  the point of view of the considered hypothesis,
in the case of a homogeneous emission of the sun 
the total radiation of the corona should be proportional to
the square of the solar constant, but for inhomogeneous radiation 
be proportional to an even higher power of it. This conclusion also
agrees with observations, as can be seen from table 2,
taken from the work of\ $\,$P$\,$e$\,$t$\,$t$\,$i$\,$t$\,$\ and 
$\,$N$\,$i$\,$c$\,$h$\,$o$\,$l$\,$s$\,$o$\,$n\footnote{E.\ 
$\,$P$\,$e$\,$t$\,$t$\,$i$\,$t$\,$\ and S$\,$e$\,$t$\,$h$\,$ B.\
$\,$N$\,$i$\,$c$\,$h$\,$o$\,$l$\,$s$\,$o$\,$n, loc.\ cit.}:
the variation of the total emission is considerably larger than
the variation of the solar constant.

\begin{center}
T$\,$A$\,$B$\,$L$\,$E$\,$\ 2.

\vspace{2mm}

\begin{tabular}{|c|c|c|}\hline\hline
& &  \\[-2mm]
Eclipse & Energy of the corona  & Solar constant\\
& & \\[-2mm]
\hline\hline
 & & \\
1918 & 
$228\cdot 10^{-8}\ \frac{\displaystyle\rm cal.}{\displaystyle\rm min.}\ 
{\it cm}^2$
&$1,94\ \frac{\displaystyle\rm cal.}{\displaystyle\rm min.}\ 
{\it cm}^2$ \\[3mm]
1922 & 165\hspace{31mm}\ & 1,89\hspace{20mm}\ \\[1mm]
1925 & 214\hspace{31mm}\ & 1,93\hspace{20mm}\ \\
 & &
\end{tabular}

\end{center}

The ``selfscattering'' hypothesis could furthermore explain the
remarkable lack of\ 
$\,$F$\,$r$\,$a$\,$u$\,$n$\,$h$\,$o$\,$f$\,$e$\,$r$\,$\ lines in
the corona layers nearest to the sun. The vanishing of these lines
might be effected by the\ $\,$C$\,$o$\,$m$\,$p$\,$t$\,$o$\,$n$\,$\ effect
during ``collisions'' of light quanta thanks to which the details
of the spectrum should be washed out and vanish practically.

It is without any doubt, that a significant part of the coronal 
phenomena, the radiant structure, the line spectrum, and others 
cannot have any relation
to ``selfscattering''. These phenomena should be explained by the 
influence of matter surrounding the sun.

The arguments in favour of the ``selfscattering'' hypothesis
given in this paragraph are of course not decisive. It is perfectly 
possible that other astrophysical\linebreak

\vspace{-0.55cm}

\setcounter{footnote}{0}
\hspace{15.61cm}{\it 563}\nolinebreak\hspace{-16.43cm} 
theories will give a better justified
explanation of the same facts and then one can raise the degree
of accuracy by which the superposition principle can be stated
and also diminish the hypothetical ``impenetrable'' extension of light
quanta.

\vspace{5mm}

{\small
\ \ \ \ \ \ \ \ \ \ \ \ \ Moscow.

Institute of Biological Physics.

\ \ \ Received at the Redaction

\ \ \ \ \ \ \ on November 14, 1928.}

\vspace{1cm}
\begin{center}
------------------

\vspace{1cm}

REMARKS ON THE EMPIRICAL ACCURACY OF THE OPTICAL SUPERPOSITION PRINCIPLE.

\vspace{2mm}

{\it By S.\ I.\ Vavilov}

\vspace{2mm}

S$\,$u$\,$m$\,$m$\,$a$\,$r$\,$y.

\end{center}

1.\ The present theory of light leaves the question open yet, 
if the superposition principle can be assumed to be of 
unlimited validity, or, if in the case of very large radiation 
densities also a peculiar scattering of colliding 
incoherent bundles, as a result of 
 ``collisions'' of light quanta, or the diffraction of light
waves off other quanta, is to be expected.

\label{563sum}
2.\ For estimating the empirical upper validity limit of the
superposition principle the situation in the vicinity of the
sun's surface is most advantageous. As the total intensity of 
the solar corona shows, in collisions of light rays near the sun
at maximum $1.8 \cdot 10^{-17}$ of the energy per {\it cm} can
be scattered. If one could attribute the total radiation of the corona
to ``selfscattering'' of light, the superposition principle near
the sun still would be fulfilled with an accuracy of $1.8 \cdot 10^{-17}$.

3.\ Up to now there is no satisfactory theory of the solar corona,
consequently preliminarily also the hypothesis of ``selfscattering''
of light near the sun can be discussed. Some properties of the 
corona, the gradient of the emission, the large variation of the
total intensity of the corona, the small variation of the solar
constant, etc., can be interpreted, at least qualitatively, 
from the point of view of the hypothesis.

\refstepcounter{ppage}

\newpage
\refstepcounter{ppage}

\renewcommand{\thepage}{\theppage}
\chead{\rm - \theppage\ -}

\phantomsection
\addcontentsline{toc}{section}{Supplementary information}
{\Large\bf Supplementary information}\\[0.3cm]

\phantomsection
\addcontentsline{toc}{subsection}{Translator's notes}
{\large\bf Translator's notes}:

\begin{itemize}
\item[1.]
Given the copyright situation in the Soviet Union at 
the time of publication (1928) of the original article it seems reasonable to
assume that the article is in the public domain.

\item[2.]
The numbers on the right margin of the English translation 
refer to the page numbers of the original
article. A facsimile of the original article is
appended (see p.\ \pageref{pagefac}).

\item[3.]
Concerning footnote \ref{555footnote2} on p.\ 555 [\pageref{555footnote2}]:
The translation is
quoted here after the English translation of {\it Trait\'e de la Lumi\`ere}:
\cite{1912huyg}, pp.\ 21-22.

\item[4.]
Concerning footnote \ref{555footnote3} on p.\ 555 [\pageref{555footnote2}]:
The translation is quoted here after: \cite{1970lomo}, specifically p.\ 251.

\item[5.]
The content of the present article constitutes an essential part
of part 2, chap.\ 1, \S\ 2 of the book 
{\it
\begin{otherlanguage}{russian}
Микроструктура света (Исследования и очерки)
\end{otherlanguage}
%{\cyrit Mikrostruktura sveta (Issledovaniya i ocherki)}
[Mikrostruktura sveta (Issledovaniya i ocherki)]/[The 
Microstructure of Light (Research and Essays)]} by Vavilov
\cite{1950vavilov}.

\end{itemize}

\vspace{0.5cm}

\phantomsection
\addcontentsline{toc}{subsection}{Noticed misprints}
{\large\bf Noticed misprints}\\

In the original article, the following misprints have been noticed and 
corrected in the English translation:

\begin{itemize}
\item[-]
Footnote \ref{555footnote3} on p.\ 555 [\pageref{555footnote2}]:
The footnote should read correctly:

\begin{otherlanguage}{russian}
М.\ В.\ Ломоносов, Слово о происхождении света. Полное 
собрание сочинений, ч.\
\end{otherlanguage}
%{\cyrrm 
%M.\ V.\ Lomonosov, Slovo o proiskhozhdenii sveta. Polnoe
%sobranie sochineni\u i, ch}.\ 
III, 1803, 
\begin{otherlanguage}{russian}
стр.\ 105.
\end{otherlanguage}
%{\cyrrm str.\ 105.}
[M.\ V.\ Lomonosov, Slovo o proiskhozhdenii sveta. Polnoe sobranie 
sochineni\u{\i}, ch.\ III, 1803, str.\ 105].

\item[-]
Footnote \ref{558footnote1} on p.\ 558 [\pageref{558footnote1}]: 
The journal page number should read correctly: 740.

\item[-]
Footnote \ref{559footnote2} on p.\ 559 [\pageref{559footnote2}]: 
The journal page number should read correctly: 202.

\item[-]
Item 2. of the Summary on p.\ 563 [\pageref{563sum}]: 
In the second sentence, {\it sm}
should read correctly: {\it cm}.

\end{itemize}

\vspace{0.5cm}

\phantomsection
\addcontentsline{toc}{subsection}{Acknowledgements}
{\large\bf Acknowledgements}\\

I am indebted to T.\ B.\ Avrutskaya (museum cabinet of academician
N.\ I.\ Vavilov, Vavilov Inst.\ General Genetics, Moscow)
and M.\ S.\ Aksent'eva (redaction of the journal Usp.\ Fiz.\ Nauk, Moscow)
for their invaluable help in exploring the copyright situation of
the original Russian article.\\
\refstepcounter{ppage}

\newpage
\refstepcounter{ppage}
\label{literature}

\phantomsection
\addcontentsline{toc}{subsection}{Literature}
{\large\bf Literature}\\[0.3cm]

{\rm\small 
[For references in Cyrillic letters, 
we apply the (new) {\it Mathematical Reviews} transliteration
(transcription) scheme
(to be found at the end of index issues of {\it Mathematical Reviews}).
Chinese references are transcribed using Hanyu Pinyin with tone marks.]
}\\[0.3cm]

\phantomsection
\addcontentsline{toc}{subsubsection}{Full details of the cited literature}
{\bf Full details of the cited literature} (ordered
alphabetically by author names):\\[-0.3cm]

%%CITATION = ZEPYA,28,299;%%
W.\ Anderson: \"Uber die Existenzm\"oglichkeit  
von kosmischem Staube in der Sonnenkorona
[On the possible existence of cosmic dust in the solar corona].
{\it Zeitschrift f\"ur Physik} {\bf 28}:1(1924)299-324
(\href{http://dx.doi.org/10.1007/BF01327186}{DOI: 
10.1007/BF01327186}). [in German]\\

%%CITATION = ZEPYA,33,273;%%
W.\ Anderson: Die physikalische Natur der Sonnenkorona. I
[The physical nature of the solar corona. I].
{\it Zeitschrift f\"ur Physik} {\bf 33}:1(1925)273-301\hfill\ \linebreak
(\href{http://dx.doi.org/10.1007/BF01328312}{DOI: 
10.1007/BF01328312}). [in German]\\

%%CITATION = ZEPYA,34,453;%%
W.\ Anderson: Die physikalische Natur der Sonnenkorona. II
[The physical nature of the solar corona. II].
{\it Zeitschrift f\"ur Physik} {\bf 34}:1(1925)453-473\hfill\ \linebreak
(\href{http://dx.doi.org/10.1007/BF01328489}{DOI: 
10.1007/BF01328489}). [in German]\\

%%CITATION = ZEPYA,35,757;%%
W.\ Anderson: Die physikalische Natur der Sonnenkorona. III
[The physical nature of the solar corona. III].
{\it Zeitschrift f\"ur Physik} {\bf 35}:10(1925)757-775\hfill\ \linebreak
(\href{http://dx.doi.org/10.1007/BF01386043}{DOI: 
10.1007/BF01386043}). [in German]\\ 

%%CITATION = ZEPYA,37,342;%%
W.\ Anderson: Die physikalische Natur der Sonnenkorona. IV
[The physical nature of the solar corona. IV].
{\it Zeitschrift f\"ur Physik} {\bf 37}:4-5(1926)342-366\hfill\ \linebreak
(\href{http://dx.doi.org/10.1007/BF01397106}{DOI: 
10.1007/BF01397106}). [in German]\\

%%CITATION = MNRAA,88,740;%%
I.\ Balanovsky, E.\ Perepelkin:
The distribution of brightness in the solar corona of 1927 June 29.
{\it Monthly Notices of the Royal Astronomical Society} 
{\bf 88}:9(1928)740-750.
The article is freely available online from the 
SAO/NASA Astrophysics Data System (ADS) site:
\href{http://articles.adsabs.harvard.edu/full/1928MNRAS..88..740B}{http://articles.adsabs.harvard.edu/full/1928MNRAS..88..740B}.\\

%%CITATION = 00410,20,1;%%
Ch.\ Huyghens:
{\it Abhandlung \"uber das Licht. Worin die Ursachen 
der Vorg\"ange bei seiner Zur\"uckwerfung und Brechung und 
besonders bei der eigenth\"umlichen Brechung des 
isl\"andischen Spathes dargelegt sind}.
Ostwald's Klassiker der Exakten Wissenschaften, Vol.\ 20. 
Verlag von Wilhelm Engelmann, Leipzig, 1890
(The book is freely available online at the Internet Archive site:
\hfill\ \linebreak
\href{http://www.archive.org/details/abhandlungberda00mewegoog}{http://www.archive.org/details/abhandlungberda00mewegoog}). 
[in German] \hfill Translated from the French original:
Ch.\ Huygens: {\it Trait\'e de la Lumi\`ere
o\`u sont explique\'es les causes de ce qui luy 
arrive dans la refexion, \& dans la refraction. Et particulierement
dans l'etrange refraction du cristal d'Islande}.
Pierre van der Aa, Leiden, 1690\hfill\ \linebreak
(The book is freely available online
from the ETH Zurich Library site given by the 
\href{http://dx.doi.org/10.3931/e-rara-3766}{DOI: 10.3931/e-rara-3766} .).
Reprinted in:
Ch.\ Huygens: {\it \OE uvres Com\-pl\`etes. Tome XIX. 
M\'ecanique Th\'eorique et Physique de 1666 \`a 1695}. 
Martinus Nijhoff, The Hague, 1937, pp.\ 451-548
(The text if freely available online from the 
Digitale Biblioteek voor de Nederlandse Letteren
(Digital Library of Dutch Literature) site:
\href{http://www.dbnl.org/tekst/huyg003oeuv19_01/downloads.php}{http://www.dbnl.org/tekst/huyg003oeuv19\_01/downloads.ph}
. The book is freely available online at the Gallica Site of the 
Biblioth\`eque Nationale de France:
\href{http://gallica.bnf.fr/ark:/12148/bpt6k77868x}{http://gallica.bnf.fr/ark:/12148/bpt6k77868x}).
English translation: \cite{1912huyg}.\\

\refstepcounter{ppage}
%%CITATION = NONE;%%
\begin{otherlanguage}{russian}
М.\ В.\ Ломоносов
\end{otherlanguage}
%{\cyrrm M.\ V.\ Lomonosov} 
[M.\ V.\ Lomonosov]:
{\it \begin{otherlanguage}{russian}
Слово о происхожденiи свѣта, новую теорiю о 
цв\'{ѣ}тахъ представляющее, 
въ публичномъ собранiи императорской Академiи наукъ iюля 1 дня
1756 года говоренное Михайломъ Ломоносовымъ
\end{otherlanguage}
%{\cyrit{Slovo o proiskhozhden\={\i}i sv\cyryat ta, novuyu teor\={\i}yu o
%tsv}}\'\ \hspace{-3mm}{\cyrit{\cyryat takh\cdprime\ predstavlyayushchee,
%v\cdprime\ publichnom\cdprime\ sobran\={\i}i imperatorsko\u i 
%Akadem\={\i}i nauk\cdprime\ \={\i}yulya 1 dnya
%1756 goda govorennoe Mikha\u ilom\cdprime\ Lomonosovym\cdprime }}
[Slovo o proiskhozhden\={\i}i sv\v{e}ta, novuyu teor\={\i}yu o
tsv\'{\v{e}}takh''\ predstavlyayushchee,
v''\ publichnom''\ sobran\={\i}i imperatorsko\u\i\ Akadem\={\i}i 
nauk''\ \={\i}yulya 1 dnya
1756 goda go\-vorennoe Mikha\u\i lom''\ Lomonosovym'']}.
\begin{otherlanguage}{russian}
Императорская Академiя наук
\end{otherlanguage}
%{\cyrrm Imperatorskaya Akadem\={\i}ya nauk\cdprime} 
[Impera\-torskaya Akadem\={\i}ya nauk''], 
St.\ Petersburg, 1757. [in Russian]
Reprinted in: 1.
{\it 
\begin{otherlanguage}{russian}
Полное собранiе сочиненiй Михайла Васильевича
Ломоносова. Часть третїя
\end{otherlanguage}
%{\cyrit{Polnoe sobran\={\i}e 
%sochinen\={\i}\u i Mikha\u ila Vasil\cprime evicha
%Lomonosova. Chast\cprime\ tret\"{\i}ya}}
[Polnoe sobran\={\i}e sochinen\={\i}\u\i\ Mikha\u\i la 
Vasil'evicha Lomonosova. Chast' tret\"{\i}ya]}.
\begin{otherlanguage}{russian}
Императорская Академiя наукъ
\end{otherlanguage}
%{\cyrrm Imperatorskaya Akadem\={\i}ya nauk\cdprime} 
[Imperatorskaya Akadem\={\i}ya nauk''], 
St.\ Petersburg, 1803, pp.\ 105-142 (The book is freely available 
online at the site:\hfill\ \linebreak
\href{http://imwerden.de/cat/modules.php?name=books&pa=showbook&pid=2666}{http://imwerden.de/cat/modules.php?name=books{\&}pa=showbook{\&}pid=2666}).\hfill\ \linebreak
2. {\it
\begin{otherlanguage}{russian}
Сочиненiя М.\ В.\ Ломоносова съ
объяснительными примѣчанiями
академика М.\ И.\ Сухомлинова
\end{otherlanguage}
%{\cyrit Sochinen\={\i}ya M.\ V.\ Lomonosova s\cdprime\ 
%ob\cdprime yasnitel\cprime nymi prim\cyryat chan\={\i}yami 
%akade\-mika M.\ I.\ Sukhomlinova} 
[Sochinen\={\i}ya M.\ V.\ Lomonosova 
s''\ ob''yasnitel'nymi prim\v{e}chan\={\i}yami 
akademika M.\ I.\ Sukhomlinova]}. Vol.\ 4.
\begin{otherlanguage}{russian}
Императорская Академiя наукъ
\end{otherlanguage}
%{\cyrrm Imperatorskaya Akadem\={\i}ya nauk\cdprime}
[Imperatorskaya Akadem\={\i}ya nauk''], St.\ Petersburg, 1898,
item XIII, p.\ 392-424
(The book is freely available online at the following site of the
Russian Academy of Sciences:\hspace{0.2cm}
{\tt \url{http://nasledie.enip.ras.ru/ras/view/publication/general.html?id=42070256}}).
3. \begin{otherlanguage}{russian}
М.\ В.\ Ломоносов
\end{otherlanguage}
%{\cyrrm M.\ V.\ Lomonosov} 
 [M.\ V.\ Lomonosov]:
\begin{otherlanguage}{russian}
Слово о происхождении света, новую теорию о 
цв\'{e}тах представляющее, 
в публичном собрании императорской Академии Наук июля 1 дня
1756 года говоренное Михайлом Ломоносовым
\end{otherlanguage}
%{\cyrrm{Slovo o proiskhozhdenii sveta, novuyu teoriyu o
%tsv}}\'\ \hspace{-3mm}{\cyrrm{etakh predstavlyayushchee,
%v publichnom sobranii imperatorsko\u i Akademii Nauk iyulya 1 dnya
%1756 goda govorennoe Mikha\u ilom Lomonosovym}}
[Slovo o proiskhozhdenii sveta, novuyu teoriyu o
tsv\'{e}takh predstavlyayushchee,
v publichnom sobranii imperatorsko\u\i\ Akademii Nauk iyulya 1 dnya
1756 goda govorennoe Mikha\u\i lom Lomonosovym].
In: {\it 
\begin{otherlanguage}{russian}
Полное собрание сочинений. Том Третий. Труды по  физике 1753-1765 гг.
\end{otherlanguage}
%{\cyrit{Polnoe sobranie sochineni\u i. Tom treti\u i.
%Trudy po fizike 1753-1765 gg.}}
[Polnoe sobranie sochineni\u\i. Tom treti\u\i .
Trudy po fizike 1753-1765 gg.]}.
\begin{otherlanguage}{russian}
Издательство Академии наук СССР
\end{otherlanguage}
%{\cyrrm Izdatel\cprime stvo Akademii nauk SSSR}
[Izdatel'stvo Akademii nauk SSSR], Moscow, 1952, item 15, pp.\ 315-344
(The book is freely available 
online at the site:\hfill\ \linebreak
\href{http://lomonosov300.ru/6158.htm}{http://lomonosov300.ru/6158.htm}).
English translation: \cite{1970lomo}, pp.\ 247-269.\\

%%CITATION = SPWPA,1925,83;%%
H.\ Ludendorff: Spektralphotometrische Untersuchungen \"uber die
Sonnenkorona\hfill\ \linebreak
\protect{[Spectral photometric investigations on the solar corona].}
{\it Sitzungsberichte der\hfill\ \linebreak 
Preussischen Akademie der Wissenschaften, Physikalisch-Mathematische
Klasse}\hfill\ \linebreak  (1925) No.\ 5, 83-113. [in German]\\

%%CITATION = NONE;%%
{\it M\"uller-Pouillets Lehrbuch der Physik.
Band 5: Physik der Erde und des Kosmos (einschl. Relativit\"atstheorie),
H\"alfte 2: Physik des Kosmos (einschl. Relativit\"ats\-theorie)},
11.\ ed.. Vieweg, Braunschweig, 1928. [in German]\\

%%CITATION = ASJOA,62,202;%%
E.\ Pettit, S.\ B.\ Nicholson:
Radiation measurements of the solar corona January 24, 1925.
{\it Astrophysical Journal} {\bf 62}:3(1925)202-225
(\href{http://dx.doi.org/10.1086/142926}{DOI: 10.1086/142926}).
The article is freely available online from the 
SAO/NASA Astrophysics Data System (ADS) site:
\href{http://articles.adsabs.harvard.edu/full/1925ApJ....62..202P}
{http://articles.adsabs.harvard.edu/full/1925ApJ....62..202P}.

\refstepcounter{ppage}

\pagebreak
\refstepcounter{ppage}

\phantomsection
\addcontentsline{toc}{subsubsection}{Further references}
\def\refname{\normalsize Further references:}

\refstepcounter{ppage}

\pagebreak
\refstepcounter{ppage}
\label{pagefac}

\refstepcounter{section}
\addcontentsline{toc}{section}{Facsimile of the original article}

\chead{\ \vspace{1cm}\sf [Facsimile of
Zh.\ Russk.\ Fiz.-Khim.\ Obshch., Ch.\ Fiz.\ {\bf 60}(1928)555-563]\hspace{0.6cm}\ }

\includepdf[pages=1-1,lastpage=9,scale=0.8,offset=0.3cm -1.5cm,
pagecommand={\pagestyle{fancy}}]{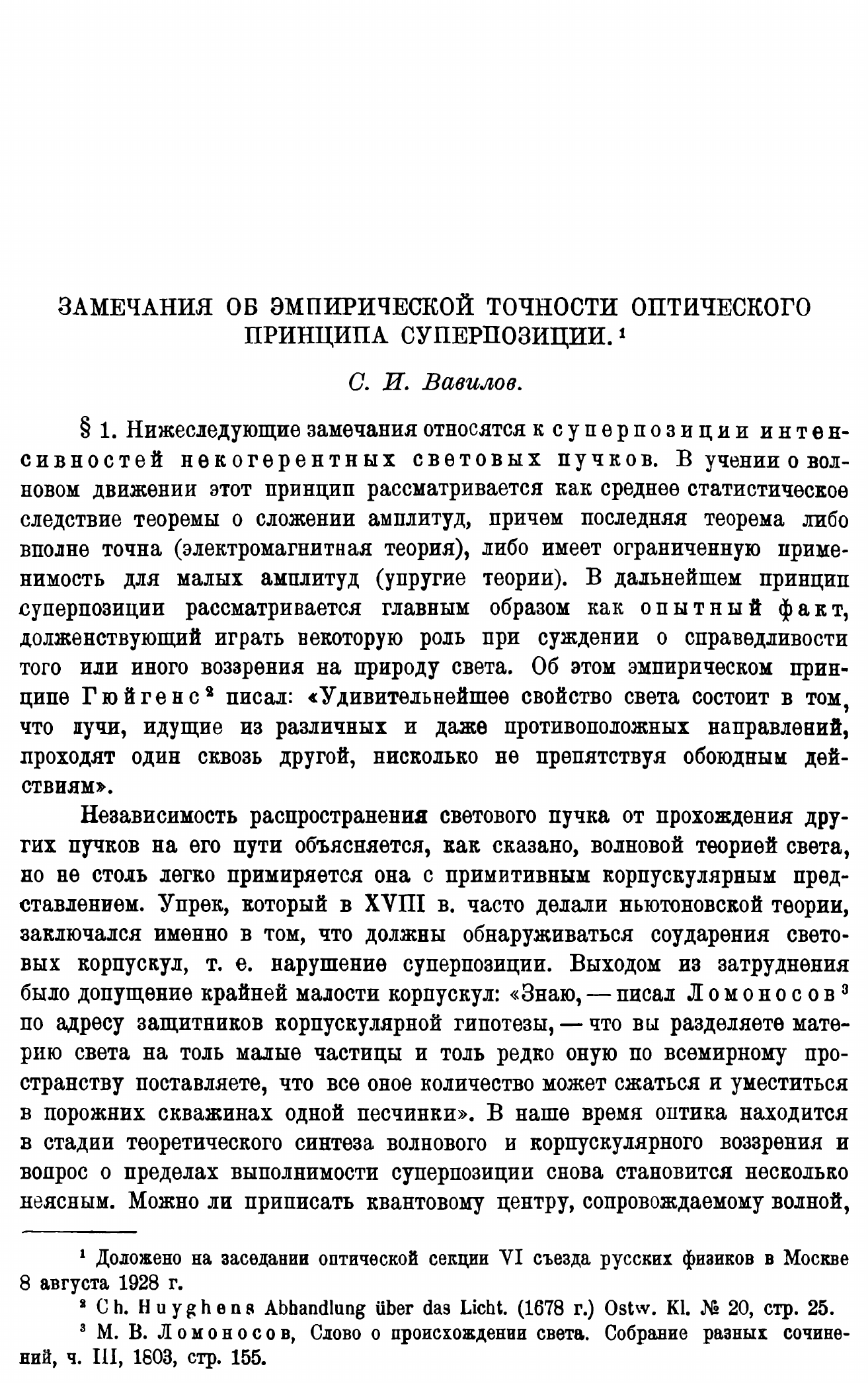}
%pagecommand={\thispagestyle{empty}}]{zhrusskfizchimobchfiz60p555dpi600bwocr.pdf}

\pagebreak

\renewcommand{\thepage}{\arabic{page}}



\section*{Supplementary material}

Общее свечение позволяет, однако, установить **верхний предел** справедливости принципа.

Для приближенного подсчета этого предела можно исходить, из того факта, что суммарное свечение короны, как показывают фотографии примерно на половину исчерпывается на расстояниях в половину солнечного радиуса от края Солнца. Для более точного расчета можно воспользоваться эмпирическими формулами для закона убывания яркости короны с расстоянием. Например, по данным Балановской и Перепелкиной[1] для затмения 29 июня 1927 г. яркость короны $I$ на расстоянии $R$ от центра Солнца ($R$ — выражено в солнечных радиусах) определяется формулой:

$$I = \frac{A_1}{(R-0{,}849)^3} \quad \ldots \ldots \ldots \ldots \ldots \ldots (1)$$

Отсюда легко вычислить, что половина общего свечения короны сосредоточена в пределах 0,4 солнечного радиуса. Суммарное свечение короны составляет приблизительно $1 \cdot 10^{-6}$ общей солнечной радиации.[2] Половина этого количества, т. е. $5 \cdot 10^{-7}$ рассеивается на протяжении в 0 4 солнечного радиуса, т. е. на пути в $2{,}8 \cdot 10^{10}$ см. Таким образом, в среднем на каждом см пути около Солнца световой луч теряет **по всем существующим там причинам рассеяния** $\frac{5 \cdot 10^{-7}}{2{,}8 \cdot 10^{10}}$, т. е. около

$$\boxed{1{,}8 \cdot 10^{-17}}$$

своей энергии. Такой верхний предел точности выполнения принципа суперпозиции вблизи Солнца, т. е. в условиях, исключительно благоприятных для его предположительного нарушения. Даже в случае отсутствия конкурирующих причин «саморассеянию» света можно было бы приписать только $1{,}8 \cdot 10^{-17}$ энергии пересекающихся весьма интенсивных световых пучков. Отсюда ясно, насколько трудно, если вообще не безнадежно, обнаружить в лаборатории такое рассеяние, если бы оно и было реальным. С конденсированной искрой легко добиться в течение короткого времени около $1 \cdot 10^{-6}$ *сек.* светового потока мощностью в $10^{11} \frac{эрг.}{см^2}$ *сек.*, такой же, как около Солнца. Светорассеяние в секунду во **все стороны** в предельном благоприятном случае могло бы составить (при 100 искрах в секунду) только $1{,}8 \cdot 10^{-17} \cdot 10^{11} \cdot 10^{-6} \cdot 10^2 = 1{,}8 \cdot 10^{-10}$ эргов в секунду. Фактически наблюдению подлежала бы величина в десятки или даже сотни раз меньшая порога зрительного раздражения.

---

[1] I. Balanovsky and E. Perepelkin. Monthly Notices of. Roy. Astr. Soc. 88, 748, 1928.

[2] Ср. Müller-Pouillet, Lehrbuch des Physik, B. V, 2-te Hälfte, стр. 157.

Слабость свечения солнечной короны показывает таким образом, что принцип суперпозиции, по крайней мере для видимого света, является одним из точнейших положений оптики.

§ 3. Можно ли, уверенно утверждая столь широкие пределы применимости суперпозиции, одновременно видеть в солнечной короне аргумент в пользу существования «саморассеяния света»? Если какой-либо астрофизической теории удается дать убедительное объяснение короны в ее наиболее интенсивной части, вопрос естественно отпадает; до тех пор пока этого нет и пока теория света не приняла конкретных форм, гипотеза «саморассеяния» как будто бы имеет право дискутироваться наравне с другими предположениями. К сожалению, прямой путь решения, земной experimentum crucis, как объяснено выше, едва ли осуществим.

Некоторые характерные свойства короны находят простое толкование с точки зрения гипотезы «саморассеяния». Яркость короны быстро убывает по мере удаления от края Солнца. Существует несколько эмпирических формул, выражающих зависимость яркости от расстояния,[1] эти формулы сильно отличаются одна от другой. Для малых расстояний от Солнца в некоторых случаях хорошо выполняется формула Тэрнера[2]

$$I = \frac{A_2}{R^6}, \quad\quad\quad\quad\quad\quad\quad (2)$$

где $R = \frac{r}{r_0}$ ($r$ — расстояние от центра Солнца, $r_0$ — радиус Солнца). Беккером предложена формула:

$$I = \frac{A_3}{(R-0,72)^4} \quad\quad\quad\quad\quad\quad (3)$$

Кубическая формула (1) приведена выше. На больших расстояниях результаты наблюдений иногда хорошо согласуются с формулой Бергстранда:

$$I = \frac{A_4}{(R-1)^2} \quad\quad\quad\quad\quad\quad (4)$$

До сих пор, повидимому, не выяснено, соответствует ли такое разнообразие формул действительным изменениям в короне, или является только результатом несовершенства измерений.

С точки зрения гипотезы о соударениях световых квантов интенсивность рассеянного света в данном месте должна быть пропорциональной числу соударений квантов, т. е. прежде всего произведению **квадрата плотности радиации на относительную скорость** соударяющихся квантов.

---

[1] Cp. W. Anderson. ZS. f. Physik 35, 757, 1926.
[2] Cp. E. Pettit und Seth. B. Nicholson. Astrophys. Journ. 62, 222, 1925.

Пусть линия $ED$ на рис. 1 соединяет рассматриваемую точку $A$ около Солнца с Землею $E$. Обозначим через $r_0$ — радиус Солнца, через $r$ — расстояние $CG$ (перпендикуляр из центра Солнца на линию $ED$), через $x$ — расстояние $AG$. Плотность радиации в точке $A$ пропорциональна:[1]

$$1 - \sqrt{1 - \frac{r_0^2}{x^2 + r^2}}.$$

Средняя относительная скорость квантов, соударящихся в точке $A$, пропорциональна sin угла, образуемого **средним** пучком конуса лучей, встречающихся в $A$, с осью конуса.

Отсюда средняя относительная скорость встречающихся в точке $A$ квантов пропорциональна:

$$\sqrt{\frac{r_0^2}{x^2 + r^2}}.$$

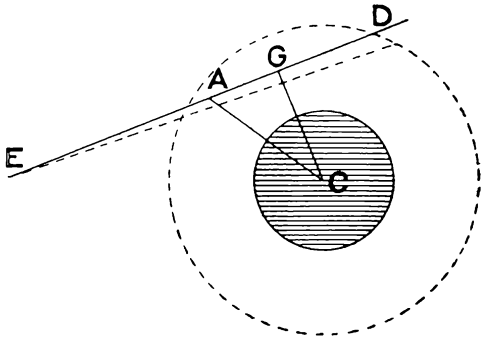

Рис. 1.

Величина и направление рассеяния для световых квантов может оказаться зависящей от угла, образуемого встречающимися квантами. Сказать что-либо об этой зависимости нельзя, поскольку само явление «соударений» квантов является совершенно гипотетическим и неопределенным. Формально мы характеризуем эту зависимость некоторой функцией

$$f(r, x).$$

Рассеяние ото всех элементарных объемов, лежащих в пределах бесконечно малого телесного угла с вершиной на Земле, складывается для земного наблюдателя в перспективе в одну точку короны; величина элементарных объемов растет пропорционально квадрату расстояния от Земли, наоборот интенсивность рассеяния, доходящая до Земли от этих объемов, убывает обратно пропорционально квадратам расстояния; таким образом, в выражение для суммарного рассеяния расстояние от Земли не войдет. Свечение в некоторой точке короны на расстоянии $r$ от центра Солнца согласно сделанным предположениям выразится формулой:

$$I = A_3 \int_0^\infty \left[1 - \sqrt{1 - \frac{r_0^2}{x^2 + r^2}}\right]^2 \cdot \left[\sqrt{\frac{r_0^2}{x^2 + r^2}}\right] \cdot f(x, r) \cdot dx. \quad \ldots \quad (5)$$

При этом сделано предположение, что радиация со всех элементов

---

[1] См. W. Anderson, loc. cit.

солнечной поверхности одинакова (что, разумеется, не точно) и, стало быть,

$$A_5 = kE^2, \quad \ldots \ldots \ldots \ldots \ldots \ldots (6)$$

где $E$ — солнечная постоянная.

Предположим, что $f(r,x)$ не зависит или мало зависит от $r$.

Ограничимся далее такими расстояниями от края Солнца, при которых среднее значение $x^2 + r^2$ под интегралом значительно больше $r_0^2$. При таких условиях формула (5) примет простой вид:

$$I = \frac{A_6}{R^4}, \quad \ldots \ldots \ldots \ldots \ldots \ldots (7)$$

где $R$ имеет прежнее значение.

Не может быть речи о точном сравнении эмпирических формул (1), (2) (3), (4) с теоретическими формулами (6) и (7). Эмпирические формулы не согласуются между собой, с другой стороны некоторая часть свечения короны несомненно вызывается рассеянием света в материи, и потому теоретическая формула не вполне применима к короне. С такими оговорками

ТАБЛИЦА 1.

| $R$ | (1) | (2) | (3) | (4) | (7) |
|---|---|---|---|---|---|
| 1,8 | 11,6 | 21,0 | 20,0 | 6,1 | 7,7 |
| 2,0 | 6,5 | 10,4 | 10,0 | 3,9 | 5,0 |
| 2,5 | 2,2 | 2,9 | 2,7 | 1,8 | 2,0 |
| 3,0 | 1,0 | 1,0 | 1,0 | 1,0 | 1,0 |
| 3,5 | 0,55 | 0,40 | 0,45 | 0,64 | 0,54 |
| 4,0 | 0,31 | 0,18 | 0,24 | 0,43 | 0,31 |
| 4,5 | 0,20 | 0,09 | 0,14 | 0,32 | 0,20 |
| 5,0 | 0,14 | 0,05 | 0,18 | 0,25 | 0,13 |

можно, однако, говорить о довольно удовлетворительном согласии теоретических и эмпирических формул. В таблице 1 сопоставлены относительные значения яркости короны от $R = 1,8$ до $R = 5,0$, вычисленные по формулам (1), (2), (3), (4) и (7), причем значения при $R = 3,0$ приравнены 1. Из таблицы видно, что формула (7) хорошо согласуется с самой свежей эмпирической формулой (1). Согласие в верхних цифрах будет лучше, если не ограничиваться приближенной формулой (7), но вычислить значения $I$ по основной формуле (5), полагая $f(x, r)$ постоянной. Сносное согласие имеется также с формулой (4), с другой стороны формулы Тэрнера (2) и Беккера (3) резко расходятся с (7).

В формулах (5) и (7), как ясно из предыдущего, постоянные $A_5$ или $A_6$ пропорциональны произведению интенсивностей встречающихся пучков. Если бы излучение солнечной поверхности всюду было одинаковым, то $A_5$ и $A_6$ были бы просто пропорциональны квадрату солнечной постоянной, что и положено в основу (5).

Вследствие **неравномерности** свечения Солнца значения постоянных $A_5$ и $A_6$ должны довольно резко отличаться по различным радиальным направлениям короны, оставаясь, однако, практически неизменными вдоль данного радиуса, что согласуется с наблюдениями [1]. Суммарное свечение короны, с точки зрения рассматриваемой гипотезы, при равномерной радиации Солнца должно быть пропорциональным квадрату солнечной постоянной, а при неравномерной радиации еще более высоким степеням. Это следствие также согласуется с наблюдениями, как видно из таблицы 2, взятой из работы Петтита и Никольсона[2]: колебания суммарного свечения значительно резче, чем колебания солнечной постоянной.

ТАБЛИЦА 2.

| Затмение | Энергия короны | Солн. постоянная |
|---|---|---|
| 1918 г. | $228 \cdot 10^{-8} \frac{\text{кал.}}{\text{мин.} \, см^2}$ | $1,94 \frac{\text{кал.}}{\text{мин.} \, см^2}$ |
| 1922 » | 165 | 1,89 |
| 1925 » | 214 | 1,93 |

Гипотеза «саморассеяния» могла бы далее объяснить примечательный факт отсутствия линий **Фраунгофера** в ближайших к Солнцу слоях короны. Исчезновение этих линий могло бы вызываться эффектом **Комптона** при «соударениях» световых квантов, благодаря чему детали спектра должны размываться и практически исчезать.

Несомненно, что значительная доля корональных явлений, лучистая структура, линейный спектр и пр. не может иметь никакого отношения к «саморассеянию». Эти явления должны объясняться действиями материи, окружающей Солнце.

Приведенные в этом параграфе некоторые доводы в пользу гипотезы «саморассеяния» света, конечно, не являются решающими. Весьма воз-

---

[1] Ср. H. Ludendorff. Sitzb. d. preuss. Akc. d. W., V, 1925, также. I. Balanovsky und E. Perepelkin, loc. cit.

[2] E. Pettit and Seth B. Nicholson, loc. cit.

можно, что другие астрофизически теории дадут более правдоподобное объяснение тех же фактов, и тогда удастся еще повысить степень точности, с которой может утверждаться принцип суперпозиции и еще понизить гипотттические «непроницаемые» размеры светового кванта.

## BEMERKUNGEN ZUR EMPIRISCHEN GENAUIGKEIT DES OPTISCHEN SUPERPOSITIONSPRINZIPS.

### Von. S. I. Wawilow.

#### Zusammenfassung.

1. Die gegenwärtge Theorie des Lichtes lässt es noch unklar, ob das Superpositionspriuzip als uubegrenzt gültig angcuommen werden darf, oder bei sehr grossen Strahlungsdichten auch eigenartige Zerstreung der zusammentreffenden inkoherenten Bündeln zu erwarten ist, infolge der «Zusammenstössen» der Lichtquanten, oder der Diffraktion der Lichtwellen an den fremden Quanten.

2. Zur Schätzung der empirischen oberen Grenze der Gültigkeit des Superpositionsprinzips sind die Verhältnisse in der Nähe der Sonnenoberfläche am vorteilhafsten. Wie die gesammte Intensität der Sonnenkorona zeigt, kann bei Zusammentreffen der Lichtstrahlen in der Nähe der Sonne grösstens nur $1,8 \cdot 10^{-17}$ der Energie pro $sm$ zerstreut werden. Wenn man die gesammte Strahlung der Korona der «Seldstxerstreuung» des Lichtes zuschreiben könnte, so würde doch das Superpositionsprinzip in der Nähe der Sonne mit der Genauihkeit $1,8 \cdot 10^{-17}$ erfüllt.

3. Bis jetzt gibt es noch keine befridiegende Theorie der Sonnenkorona, so dass vorläufig auch die Hypothese der «Seldststreuung» des Lichtes in der Nähe der Sonne diskutiert werden kunn. Einige Eigenschaften der Korona, der Gradient des Leuchlens, die grossen Schwankungen der gesammten Intensitat der Korona dei kleinen Schwankungen der Solarkonstante u. s. w. können vom Standpunkte dieser Hypothese wenigstens qualitativ gedeutet warden.



\end{document}